\documentclass[ALICE,manyauthors]{cernphprep}
\pdfoutput=1
\usepackage[comma,square,numbers,sort&compress]{natbib}
\usepackage{hyperref}
\usepackage{lineno}
\usepackage{xspace}
\usepackage{rotating}
\usepackage[T1]{fontenc}
\usepackage{multirow}
\usepackage[toc,page]{appendix}
\usepackage{graphicx}

\usepackage{tikz,colortbl}
\usetikzlibrary{calc}
\usepackage{zref-savepos}

\begin{document}

%

\newcommand{\pp}           {pp\xspace}
\newcommand{\ppbar}        {\mbox{$\mathrm {p\overline{p}}$}\xspace}
\newcommand{\XeXe}         {\mbox{Xe--Xe}\xspace}
\newcommand{\PbPb}         {\mbox{Pb--Pb}\xspace}
\newcommand{\pA}           {\mbox{pA}\xspace}
\newcommand{\pPb}          {\mbox{p--Pb}\xspace}
\newcommand{\AuAu}         {\mbox{Au--Au}\xspace}
\newcommand{\dAu}          {\mbox{d--Au}\xspace}

\newcommand{\s}            {\ensuremath{\sqrt{s}}\xspace}
\newcommand{\snn}          {\ensuremath{\sqrt{s_{\mathrm{NN}}}}\xspace}
\newcommand{\pt}           {\ensuremath{p_{\rm T}}\xspace}
\newcommand{\meanpt}       {$\langle p_{\mathrm{T}}\rangle$\xspace}
\newcommand{\ycms}         {\ensuremath{y_{\rm CMS}}\xspace}
\newcommand{\ylab}         {\ensuremath{y_{\rm lab}}\xspace}
\newcommand{\etarange}[1]  {\mbox{$\left | \eta \right |~<~#1$}}
\newcommand{\yrange}[1]    {\mbox{$\left | y \right |~<~#1$}}
\newcommand{\dndy}         {\ensuremath{\mathrm{d}N_\mathrm{ch}/\mathrm{d}y}\xspace}
\newcommand{\dndeta}       {\ensuremath{\mathrm{d}N_\mathrm{ch}/\mathrm{d}\eta}\xspace}
\newcommand{\avdndeta}     {\ensuremath{\langle\dndeta\rangle}\xspace}
\newcommand{\dNdy}         {\ensuremath{\mathrm{d}N_\mathrm{ch}/\mathrm{d}y}\xspace}
\newcommand{\Npart}        {\ensuremath{N_\mathrm{part}}\xspace}
\newcommand{\meanNpart}        {\ensuremath{\langle N_\mathrm{part} \rangle}\xspace}
\newcommand{\Ncoll}        {\ensuremath{N_\mathrm{coll}}\xspace}
\newcommand{\dEdx}         {\ensuremath{\textrm{d}E/\textrm{d}x}\xspace}
\newcommand{\RpPb}         {\ensuremath{R_{\rm pPb}}\xspace}
\newcommand{\raa}         {\ensuremath{R_{\mathrm{AA}}}\xspace}
\newcommand{\TAA} {\ensuremath{\langle T_{\mathrm{AA}} \rangle}\xspace}
\newcommand{\Axe} {\ensuremath{A \times \epsilon}\xspace}
\newcommand{\Fnorm} {\ensuremath{F_{\mathrm{ norm}}}\xspace}

\newcommand{\nineH}        {$\sqrt{s}~=~0.9$~Te\kern-.1emV\xspace}
\newcommand{\seven}        {$\sqrt{s}~=~7$~Te\kern-.1emV\xspace}
\newcommand{\twoH}         {$\sqrt{s}~=~0.2$~Te\kern-.1emV\xspace}
\newcommand{\twosevensix}  {$\sqrt{s}~=~2.76$~Te\kern-.1emV\xspace}
\newcommand{\five}         {$\sqrt{s}=5.02$~Te\kern-.1emV\xspace}
\newcommand{\twosevensixnn}{$\sqrt{s_{\mathrm{NN}}}=2.76$~Te\kern-.1emV\xspace}
\newcommand{\fivenn}       {$\sqrt{s_{\mathrm{NN}}}=5.02$~Te\kern-.1emV\xspace}
\newcommand{\LT}           {L{\'e}vy-Tsallis\xspace}
\newcommand{\GeVc}         {Ge\kern-.1emV/$c$\xspace}
\newcommand{\MeVc}         {Me\kern-.1emV/$c$\xspace}
\newcommand{\TeV}          {Te\kern-.1emV\xspace}
\newcommand{\GeV}          {Ge\kern-.1emV\xspace}
\newcommand{\MeV}          {Me\kern-.1emV\xspace}
\newcommand{\GeVmass}      {Ge\kern-.1emV/$c^2$\xspace}
\newcommand{\MeVmass}      {Me\kern-.2emV/$c^2$\xspace}
\newcommand{\lumi}         {\ensuremath{\mathcal{L}}\xspace}

\newcommand{\ITS}          {\rm{ITS}\xspace}
\newcommand{\TOF}          {\rm{TOF}\xspace}
\newcommand{\ZDC}          {\rm{ZDC}\xspace}
\newcommand{\ZDCs}         {\rm{ZDCs}\xspace}
\newcommand{\ZNA}          {\rm{ZNA}\xspace}
\newcommand{\ZNC}          {\rm{ZNC}\xspace}
\newcommand{\SPD}          {\rm{SPD}\xspace}
\newcommand{\SDD}          {\rm{SDD}\xspace}
\newcommand{\SSD}          {\rm{SSD}\xspace}
\newcommand{\TPC}          {\rm{TPC}\xspace}
\newcommand{\TRD}          {\rm{TRD}\xspace}
\newcommand{\VZERO}        {\rm{V0}\xspace}
\newcommand{\VZEROA}       {\rm{V0A}\xspace}
\newcommand{\VZEROC}       {\rm{V0C}\xspace}
\newcommand{\Vdecay} 	   {\ensuremath{V^{0}}\xspace}

\newcommand{\ee}           {\ensuremath{e^{+}e^{-}}} 
\newcommand{\pip}          {\ensuremath{\pi^{+}}\xspace}
\newcommand{\pim}          {\ensuremath{\pi^{-}}\xspace}
\newcommand{\kap}          {\ensuremath{\rm{K}^{+}}\xspace}
\newcommand{\kam}          {\ensuremath{\rm{K}^{-}}\xspace}
\newcommand{\pbar}         {\ensuremath{\rm\overline{p}}\xspace}
\newcommand{\kzero}        {\ensuremath{{\rm K}^{0}_{\rm{S}}}\xspace}
\newcommand{\lmb}          {\ensuremath{\Lambda}\xspace}
\newcommand{\almb}         {\ensuremath{\overline{\Lambda}}\xspace}
\newcommand{\Om}           {\ensuremath{\Omega^-}\xspace}
\newcommand{\Mo}           {\ensuremath{\overline{\Omega}^+}\xspace}
\newcommand{\X}            {\ensuremath{\Xi^-}\xspace}
\newcommand{\Ix}           {\ensuremath{\overline{\Xi}^+}\xspace}
\newcommand{\Xis}          {\ensuremath{\Xi^{\pm}}\xspace}
\newcommand{\Oms}          {\ensuremath{\Omega^{\pm}}\xspace}
\newcommand{\degree}       {\ensuremath{^{\rm o}}\xspace}


\def\jpsi {\ensuremath{\mathrm{J}/\psi}\xspace}
\def\psitwos{\ensuremath{\psi {\rm (2S)}}\xspace}
\def\upsns{\ensuremath{\Upsilon{\rm (nS)}}\xspace}
\def\upsones{\ensuremath{\Upsilon{\rm (1S)}}\xspace}
\def\upstwos{\ensuremath{\Upsilon{\rm (2S)}}\xspace}
\def\upsthrees{\ensuremath{\Upsilon{\rm (3S)}}\xspace}
\def\sigtot {\ensuremath{\sigma_{\rm tot}}\xspace}
\def\dsigdpt {\ensuremath{\frac{d\sigma}{d\pt}}\xspace}
\def\dsigdy {\ensuremath{\frac{d\sigma}{dy}}\xspace}
\def\dsigdptdy {\ensuremath{\frac{d^2\sigma}{d\pt dy}}\xspace}
\newcommand{\rapidity}         {\ensuremath{y}\xspace}
\def\tomumu {\ensuremath{\rightarrow\mu^{+}\mu^{-}}\xspace}

\begin{titlepage}
\PHyear{2020}       
\PHnumber{207}      
\PHdate{6 November}  

\title{$\Upsilon$~production and nuclear modification at forward rapidity in Pb--Pb~collisions at $\mathbf{\sqrt{\textit{s}_{\textbf{NN}}}=5.02}$~\TeV}
\ShortTitle{$\Upsilon$ production and nuclear modification in \PbPb collisions}   

\Collaboration{ALICE Collaboration\thanks{See Appendix~\ref{app:collab} for the list of collaboration members}}
\ShortAuthor{ALICE Collaboration} 

\begin{abstract}
The production of $\Upsilon$~mesons in \PbPb~collisions at a centre-of-mass energy per nucleon pair \fivenn is measured with the muon spectrometer of the ALICE detector at the LHC. The yields as well as the nuclear modification factors are determined in the forward rapidity region $2.5<y<4.0$, as a function of rapidity, transverse momentum and collision centrality. The results show that the production of the \upsones meson is suppressed by a factor of about three with respect to the production in proton--proton collisions.
For the first time, a significant signal for the \upstwos meson is observed at forward rapidity, indicating a suppression stronger by about a factor 2--3 with respect to the ground state.
The measurements are compared with transport, hydrodynamic, comover and statistical hadronisation model calculations.

\end{abstract}
\end{titlepage}

\setcounter{page}{2} 


\section{Introduction} \label{intro}

The collisions of ultra-relativistic heavy ions are investigated to explore deconfined and chirally restored matter at high temperatures, the quark--gluon plasma~(QGP)~\cite{Braun-Munzinger:2015hba}. The characterisation of the degrees of freedom of the QGP as well as the transition from this new state of matter to ordinary hadrons are central questions to this field of research.
The production rate of bound states of two heavy quark-antiquark pair, quarkonia, was proposed as a key observable for deconfinement~\cite{Matsui:1986dk}. 
Quarkonium states are the best approximations in nature of two static colour charges, hence representing a unique test system of the strong interaction. 
Since bottom and charm quarks have masses $m_{\rm b/c}$ well above the temperature reached in heavy-ion collisions, they are produced dominantly within a short time scale of the order of $1/(2\cdot m_{\rm b/c})$ at the beginning of the collision by hard partonic interactions~\cite{BraunMunzinger:2000dv,Zhang:2008zzc}. Therefore, heavy quarks experience the whole evolution of the thermodynamic system.
As of today, the thermal properties of quarkonium are the subject of intense studies based on lattice quantum chromodynamics~(QCD) and effective field theories of QCD~\cite{Rothkopf:2019ipj}. These calculations observe strong modifications of the real and imaginary part of the potential between the heavy quark and its antiquark extracted with increasing precision~\cite{Mocsy:2013syh,Rothkopf:2019ipj}.
Beyond the thermal properties of quarkonia in QCD matter, the investigation of the full quantum dynamical treatment of the time evolution involving the interaction between heavy-quark pairs and the medium has started~\cite{Akamatsu:2020ypb}, with the ultimate goal of deriving the relevant observables from QCD first principles~\cite{Rothkopf:2019ipj}.

Experimentally, charmonium production in nucleus--nucleus collisions was investigated at the SPS, RHIC and LHC~\cite{Andronic:2015wma}.
The deviation of the production in nucleus--nucleus collisions with respect to the expectation from the proton--proton binary collision scaling is quantified via the nuclear modification factor.
At the LHC, the suppression of \jpsi~production in heavy-ion collisions is weaker than measurements at lower energies. This behaviour is commonly interpreted as a sign of (re)generation of charmonium either solely at the phase boundary~\cite{BraunMunzinger:2000px} or during the deconfined stage of the medium evolution~\cite{Thews:2000rj}. Both scenarios are advertised as signatures of deconfinement.
In the bottomonium sector, the CMS collaboration at the LHC pioneered the measurements with the observation of a strong suppression of the \upsones state in \PbPb collisions~\cite{Chatrchyan:2012np,Chatrchyan:2012lxa,Khachatryan:2016xxp,Sirunyan:2018nsz}.
Recently, the ALICE and CMS collaborations published the first  measurement of the second Fourier coefficient of the azimuthal anisotropy of the \upsones production that indicates a weaker elliptic flow than the one measured for \jpsi~\cite{Acharya:2019hlv,Sirunyan:2020qec}. The latter measurement provides a new experimental handle to which extent the emerging bound state participates in the collective motion or is sensitive to path-length dependence of the nuclear modification. The statistical precision of the present measurements is not yet sufficient for strong model constraints.
Furthermore, the production of the excited \upstwos state is much more strongly suppressed whereas the \upsthrees~state has not been observed yet and is at least similarly suppressed as the \upstwos state~\cite{Chatrchyan:2012np,Chatrchyan:2012lxa,Sirunyan:2017lzi,Khachatryan:2016xxp,Sirunyan:2018nsz}. 
The interpretation of the data requires to account for the feed-down from the decay of excited states as well as the consideration of effects not related to the occurrence of the QGP, for example, differences between nuclear parton distribution functions~(PDFs) and free nucleon PDFs.
Nuclear modification factors were also measured in proton--nucleus collisions by the LHC collaborations~\cite{Abelev:2014oea,Aaij:2014mza,Aaboud:2017cif,Aaij:2018scz,Acharya:2019lqc} where no formation of a plasma phase was expected prior to the LHC and RHIC measurements~\cite{Khachatryan:2010gv,CMS:2012qk,Abelev:2012ola,Aad:2012gla,Aaij:2015qcq,ALICE:2017jyt,PHENIX:2018lia}. 
The results indicate a significant modification of the \upsones production at midrapidity and forward rapidity, in line with expectations from nuclear PDFs~\cite{Kovarik:2015cma,Eskola:2016oht,Kusina:2017gkz} or energy loss considerations~\cite{Arleo:2012rs}. However, the experimental data hint of a stronger suppression than present in these approaches at backward rapidity. 
In addition, stronger suppressions of the excited states compared to the \upsones~state were observed pointing towards other possible mechanisms, potentially analogue to nucleus--nucleus collisions~\cite{Ferreiro:2018wbd}.

Currently, the status of phenomenology in nucleus--nucleus collisions comprises frameworks implementing transport or rate  equations~\cite{Krouppa:2016jcl,Du:2017qkv,Ferreiro:2018wbd,Yao:2020xzw, Islam:2020gdv} using a wide range of approaches for the created thermodynamic system and its interaction with the b$\overline{\rm b}$~quark pair. The statistical hadronisation model, usually applied to the charm sector, was also proposed as a possible scenario for the bottom sector including the production of bottomonia~\cite{Andronic:2017pug}.

The ALICE collaboration reported the suppression of $\Upsilon$~production at forward rapidity $2.5 < y < 4.0$ in Pb--Pb collisions at \twosevensixnn~\cite{Abelev:2014nua} and 5.02~\TeV~\cite{Acharya:2018mni} based on the 2011 and 2015 data samples, respectively. 
This article presents the combined analysis of the 2015 and 2018 data sets recorded at \fivenn, corresponding to a three times larger integrated luminosity than the previously published measurements at the same energy.
This increase enables to perform a detailed measurement of the \upsones production as a function of the centrality, transverse momentum and rapidity.
For the first time, a significant \upstwos signal is observed in the forward rapidity region in heavy-ion collisions.
\section{Detector, data samples and observables} \label{detector}

A detailed description of the ALICE setup and its performance can be found in the references~\cite{ALICEdetector,ALICEperformance}.  The analysis is based on the detection of muons within the pseudorapidity range $-4.0 < \eta < -2.5$\footnote{In the ALICE reference frame, the muon spectrometer covers a negative $\eta$ range and consequently a negative $y$ range. We have chosen to present our results with a positive $y$ notation, due to the symmetry of the collision system.}
reconstructed and identified with the muon spectrometer~\cite{Aamodt:2011gj}. In the following, we briefly introduce the detection systems relevant for the $\Upsilon$~measurements in \PbPb~collisions.

The primary vertex is determined with the silicon pixel detector, the two innermost layers of the inner tracking system of the central barrel of ALICE~\cite{ITS}. These two cylinders surrounding the beam pipe cover the pseudorapidity range $\left | \eta \right |< 2$ (first layer) and $\left | \eta \right | < 1.4$ (second layer) assuming the nominal interaction point (IP). The V0 detector~\cite{vzero} provides the centrality determination. It is made of two arrays of scintillators covering the pseudorapidity intervals $-3.7<\eta<-1.7$ and $2.8<\eta<5$. The logical AND of the signals from both subdetectors constitutes the minimum bias~(MB) interaction trigger.  This trigger decision is fully efficient for the 0--90$\%$ most central collisions. Zero-degree calorimeters~\cite{ALICE:2012aa}, installed at $\pm 112.5$~m from the IP along the beam direction, are used for the rejection of events corresponding to electromagnetic interactions of the colliding lead nuclei. 

The muon spectrometer of ALICE consists of a front absorber to filter muons, five tracking stations, a dipole magnet with a field integral of 3~T$\times$m surrounding the third tracking station, an iron wall to reject further punch-through hadrons and low momentum muons from pion and kaon decays, and two trigger stations. These elements are traversed subsequently by the muons originating from the IP region.
Each tracking station is composed of two planes of cathode-pad chambers. The two trigger stations are equipped with two planes of resistive plate chambers~\cite{Bossu:2012jt}.

The trigger used for this analysis requires a coincidence of the MB signal and a dimuon trigger provided by the trigger stations. The dimuon condition consists of a positively and a negatively charged track candidate above a transverse momentum threshold of 1~\GeVc each. The analysed data set was recorded in 2015 and 2018 and corresponds to a total integrated luminosity of about 760~$\mu$b$^{-1}$. 

The centrality determination relies on a Glauber fit to the sum of the signal amplitudes of the \VZERO~detectors~\cite{Abelev:2013qoq,Adam:2015ptt,ALICE-PUBLIC-2018-011}. The centrality ranges are quoted as quantiles in percent of the total inelastic \PbPb cross section. The fit allows each centrality interval to be attributed an average number of participants \meanNpart, of  binary nucleon--nucleon collisions $\langle \Ncoll \rangle$ and the average nuclear overlap function \TAA. The analysis comprises measurements integrated over the centrality interval 0--90$\%$ and differential in centrality. The Glauber fit quantities used in this analysis are tabulated in the note~\cite{ALICE-PUBLIC-2018-011}.
For the centrality intervals used in the present analysis and not reported in the note, the relevant values, including uncertainties, were obtained by averaging the corresponding values over the narrower ranges.

The nuclear modification factor is the main observable quantifying the difference of production between \PbPb and proton--proton~(pp)~collisions. It is defined as

\begin{equation} \label{eqn:raa}
    \raa = \frac{\text{d}^{2} N_{\Upsilon \rightarrow \mu^{+} \mu^{-}}/ \text{d}y \text{d}p_{\text{T}}} {\TAA \times \text{d}^{2} \sigma^{\mathrm{pp}}_{\Upsilon \rightarrow \mu^{+} \mu^{-}} / \text{d}y \text{d}p_{\text{T}}} \ \mathrm{with} \ \frac{\text{d}^{2} N_{\Upsilon \tomumu}}{\text{d}y \text{d}p_{\text{T}}} = \frac{N^{\text{raw}}_{\Upsilon \rightarrow \mu^{+} \mu^{-}}} {(A\times\epsilon)_{\Upsilon \rightarrow \mu^{+} \mu^{-}} \times N_{\text{MB}} \times \Delta y \Delta p_{\text{T}}},
\end{equation}

where $N^{\text{raw}}_{\Upsilon \tomumu}$ denotes the raw number of $\Upsilon$~candidates reconstructed via the dimuon decay channel within a given rapidity, centrality and transverse momentum interval, $(A\times\epsilon)_{\Upsilon \tomumu}$ the correction for the geometrical acceptance of the decay muons and the trigger, tracking and track quality selection inefficiencies. The quantity $N_{\rm MB}$ represents the number of equivalent inelastic nucleus--nucleus collisions, \TAA the average nuclear overlap function,
$\Delta y$ and $\Delta p_{\text{T}} $ the widths of the rapidity and transverse momentum intervals. The term $\text{d}^{2}\sigma^{\text{pp}}_{\Upsilon \tomumu} / \text{d}y \text{d}p_{\text{T}}$ corresponds to the product of the $\Upsilon$~branching ratio for the dimuon decay channel BR($\Upsilon\tomumu$) and the $\Upsilon$~production cross section in pp~collisions at the same collision energy and in the same kinematic regime of the \PbPb measurement. 
The yields $N_{\Upsilon \tomumu}$ normalised by \TAA and the yield ratio between the \upstwos and \upsones states are also presented as complementary observables. The latter reads
\begin{equation}
    \frac{N_{\upstwos}} {N_{\upsones}} = \frac{N^{\text{raw}}_{\upstwos \tomumu}}{N^{\text{raw}}_{\upsones \tomumu}} \times 
    \frac{\text{BR}(\upsones \tomumu) \times (\Axe)_{\upsones \tomumu}}{\text{BR}(\upstwos \tomumu) \times (\Axe)_{\upstwos \tomumu} }.
\end{equation}
In this equation, the yields are corrected for the branching ratios for comparison with model calculations. The relative nuclear modification factor is written simply as
\begin{equation} \label{eq:double_ratio}
    \raa (\upstwos/\upsones) = \frac{N^{\text{raw}}_{\upstwos \tomumu}}{N^{\text{raw}}_{\upsones \tomumu}} 
    \times
    \frac{(\Axe)_{\upsones\tomumu} }{(A\times\epsilon)_{\upstwos \tomumu}}
    \times
    \frac{\sigma^{\text{pp}}_{\upsones \tomumu}}{\sigma^{\text{pp}}_{\upstwos \tomumu}}.
\end{equation}

In the following, the expression ``integrated'' stands for ``integrated over the 0--90$\%$ centrality interval, in $2.5<y<4.0$ and for $\pt<15$~\GeVc'' unless otherwise specified.

\section{Analysis description} \label{sec:analysis}

\subsection{Signal extraction} \label{subsec:signal}

\begin{figure}
    \begin{center}
    \includegraphics[width=.49\linewidth]{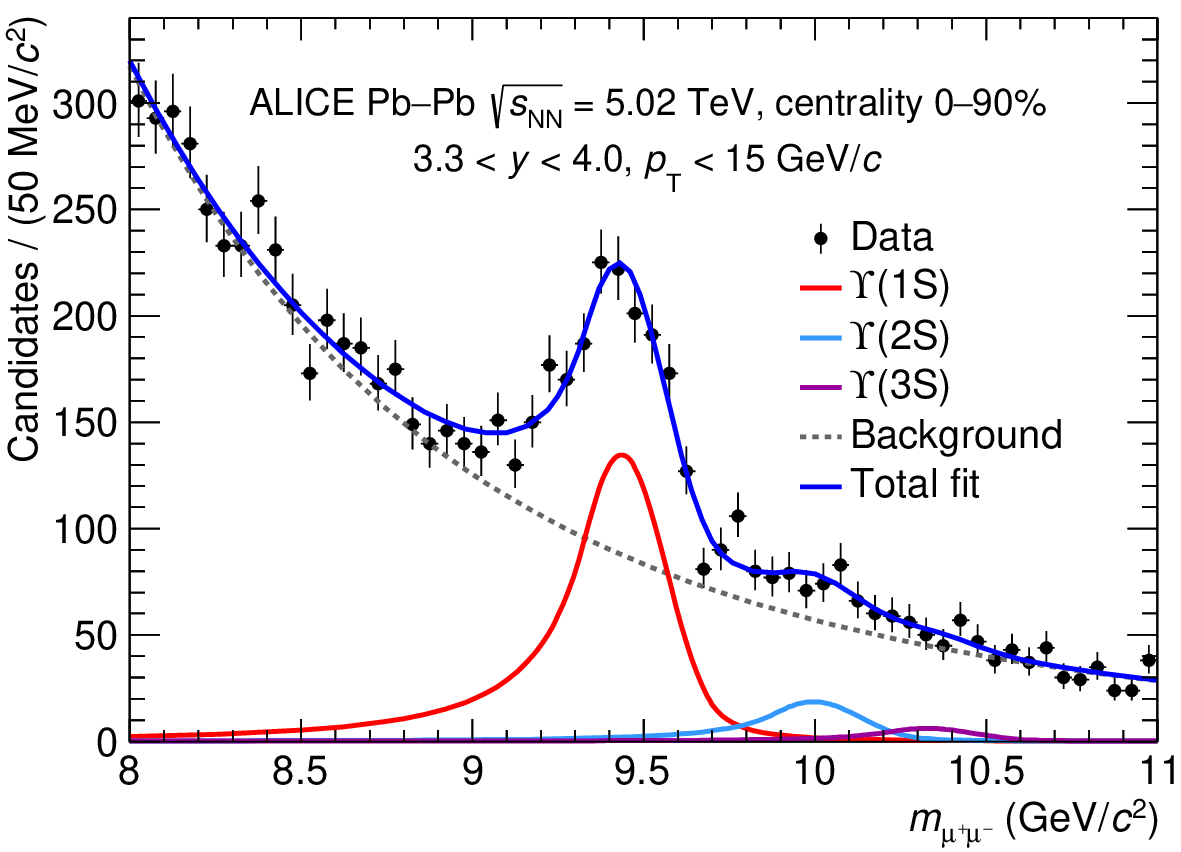}
    ~
    \includegraphics[width=.49\linewidth]{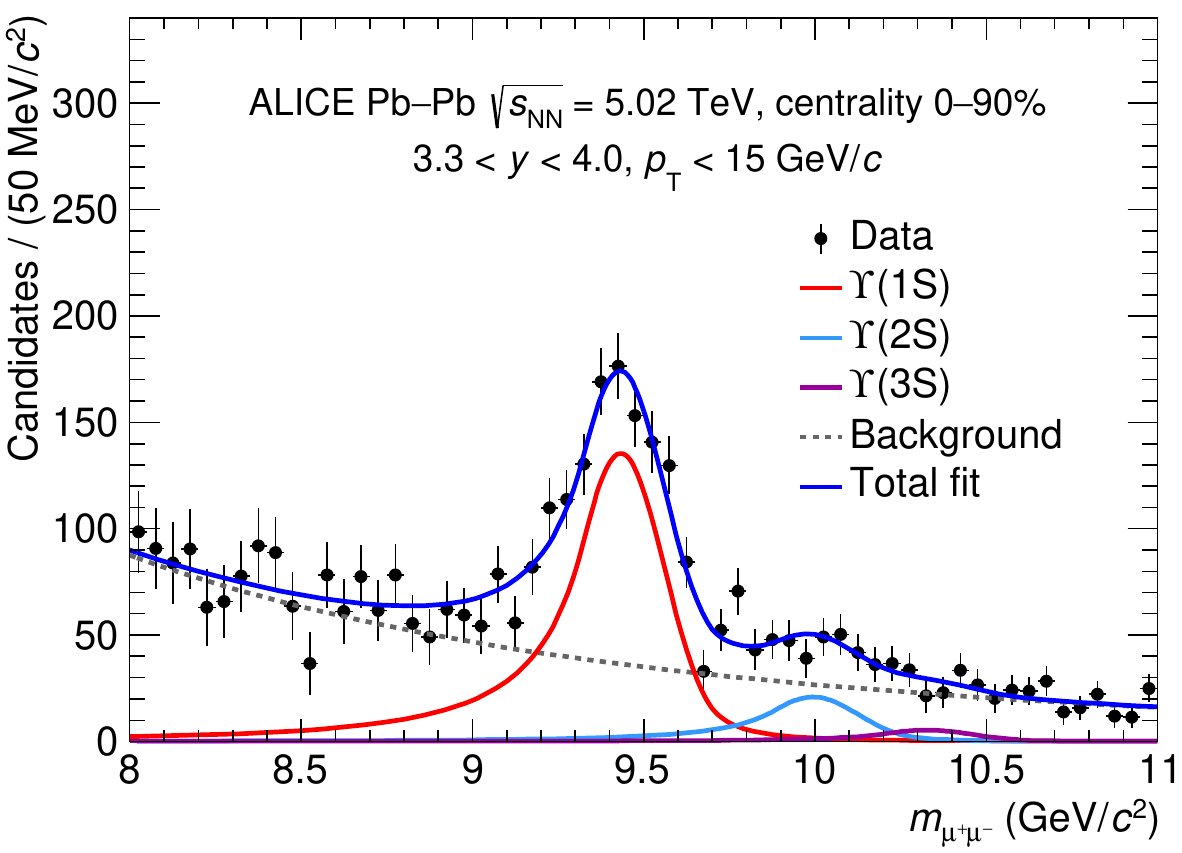}
    \caption{Invariant mass distributions of opposite-charge muon pairs within the $3.3 < y < 4.0$ interval. The left panel shows the raw invariant mass spectrum while the mixed-event background is subtracted in the right panel.}
    \label{fig:fits}
    \end{center}
\end{figure}

The number of $\Upsilon$ mesons is extracted from an unbinned log-likelihood fit to the dimuon invariant mass spectra. Opposite-charge pairs are formed with muon tracks meeting the selection criteria listed in the publication~\cite{Abelev:2014nua}. 
An extended Crystal~Ball distribution~\cite{ALICEpage} models the signal shape, one for each of the three resonances. Due to the significant background in the $\Upsilon$ mass region, the parameters quantifying the tails of the distribution have to be fixed to the values obtained from Monte~Carlo~(MC) simulations. The pole mass and width of the \upsones fit are left free. For the \upstwos and \upsthrees, the mass peak positions are fixed to the ratio of the mass values from the Particle Data Group~\cite{PDG} with respect to the \upsones while the widths are determined according to the ratio of the widths obtained in the MC simulation.
The background contribution is evaluated by empirical functions, either via a direct fit on the raw distribution or after application of the event-mixing technique.
This method, described in Ref.~\cite{eventmixing}, consists of pairing muon tracks from different events belonging to the same centrality class in order to estimate the combinatorial background present in the invariant mass distribution.
Fit examples are displayed in Fig.~\ref{fig:fits} for the two background descriptions.
After subtraction of the mixed-event background from the raw distribution, the visible residual background is also fitted with an empirical function.

In total, $N^{\text{raw}}_{\upsones\tomumu} = 3581 \pm 119$~(stat.)~$\pm\, 156$~(syst.) and $N^{\text{raw}}_{\upstwos\tomumu} = 325 \pm 61$~(stat.)~$\pm\, 60$~(syst.) are found. The quoted uncertainties are discussed in Section~\ref{subsec:syst}. No significant \upsthrees~signal is observed in any of the kinematic regions.
Assuming that the event-mixing technique provides the correct background estimation, this method improves the effective signal-to-background ratio of the \upsones by more than 2.5 and of the \upstwos by almost a factor 2 with respect to the direct extraction of the raw yields.
However, as the description of the background shape is a delicate part of this analysis, the two methods are applied for the signal extraction.

\subsection{Acceptance and efficiency corrections} \label{subsec:efficiency}

The raw yields are corrected for the acceptance and the reconstruction efficiency with a Monte~Carlo simulation. 
The $\Upsilon$~signals are generated according to \pt and $y$ distributions extrapolated from collider data~\cite{PhysRevLett.88.161802, LHCb:2012aa}, assuming an unpolarised production~\cite{Chatrchyan:2012woa,Aaij:2017egv}. These input shapes are adjusted by a nuclear PDF (nPDF) parametrisation to emulate the modification of the initial distributions.
The generated resonances are then decayed into muon pairs using the EvtGen package~\cite{evtgen}, together with \textsc{Photos}~\cite{photos} to account for final-state radiation.
Muons are transported through a realistic modelling of the ALICE apparatus via the \textsc{Geant}3 code~\cite{geant3}, and are injected in recorded MB events from data. This approach allows the experimental data-taking conditions and the occupancy of the detection elements to be accounted for. 

\begin{figure}[h]
    \begin{center}
    \includegraphics[width=.49\linewidth]{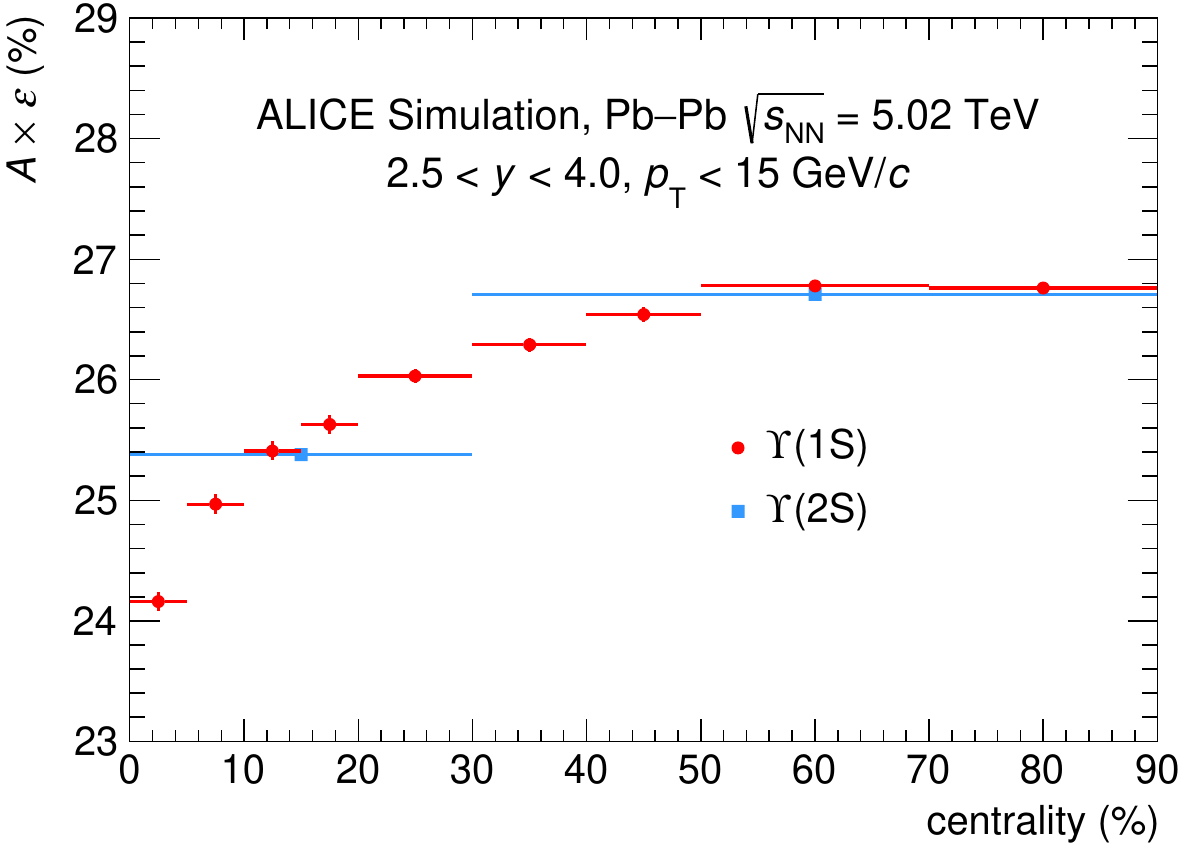}
    ~
    \includegraphics[width=.49\linewidth]{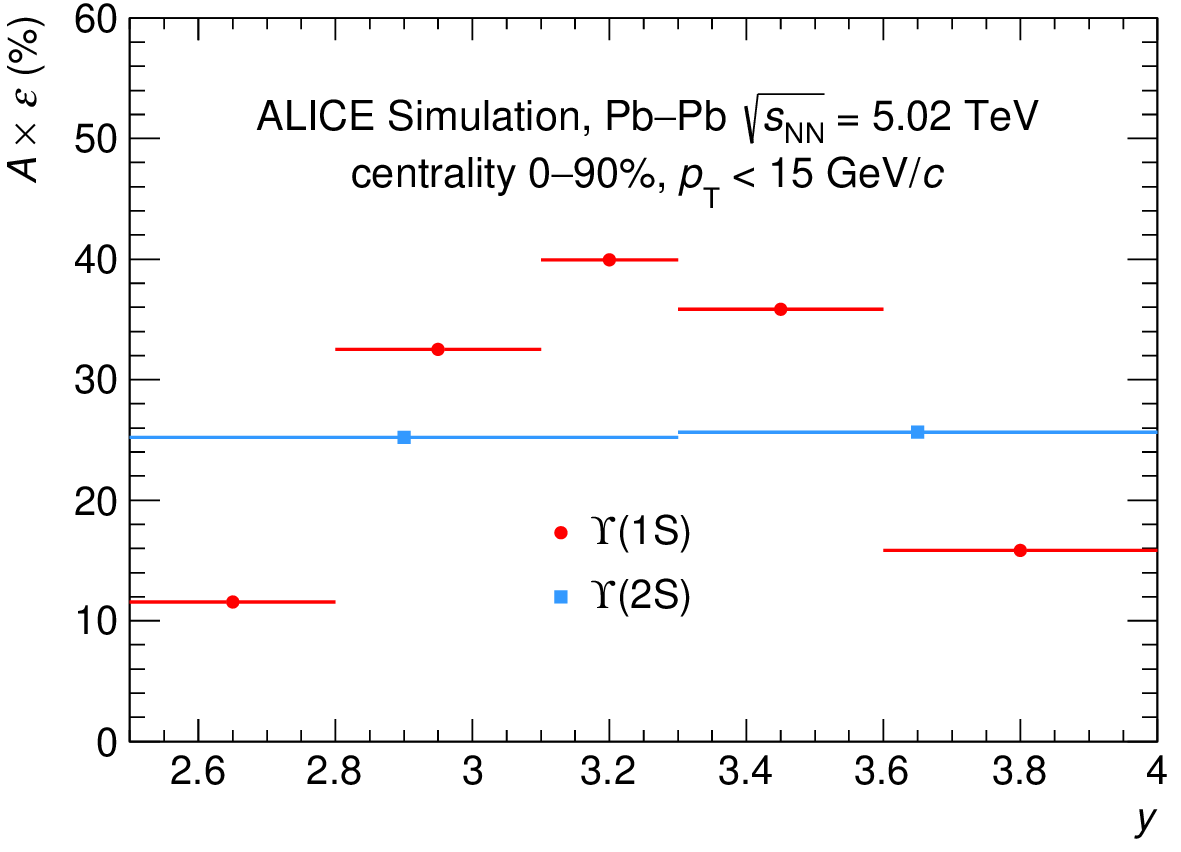}
    \caption{Acceptance and reconstruction efficiency (\Axe) for \upsones and \upstwos as a function of centrality (left) and rapidity (right). The vertical error bars denote the statistical uncertainties.}
    \label{fig:efficiency}
    \end{center}
\end{figure}

The integrated acceptance and efficiency (\Axe), averaged over the 2015 and 2018 data sets, is about $25.7\%$ for all $\Upsilon$~states. 
This value varies with \pt by less than $1\%$. Figure~\ref{fig:efficiency} shows the \Axe as a function of the collision centrality (left) and rapidity (right). An observed relative decrease of about $10\%$ from peripheral to central events is due to the rise of the occupancy in the muon chambers. The variations as a function of rapidity are a direct consequence of the spectrometer geometry.

\subsection{Production cross sections in proton--proton collisions} \label{subsec:pp}

The inclusive production of $\Upsilon$ mesons in pp collisions at \five has been recently measured at forward rapidity, $2.5 < y < 4.0$~\cite{ALICE:2021qlw}. 
The relative statistical uncertainties are of the order of the ones obtained in the present \PbPb analysis and limit the potential of the \raa study.
To overcome this limitation, the $\Upsilon$ production cross sections in pp collisions at \five are interpolated from measurements performed at various centre-of-mass energies at the LHC.

The first procedure exploits the ALICE~\cite{ALICEpp7TeV,ALICEpp8TeV} and LHCb~\cite{LHCb276TeV,LHCb78TeV,LHCb13TeV} data within the $2.5 < y < 4.0$ acceptance.
Empirical functions, such as an exponential or a power law, are fitted to the cross sections as a function of the centre-of-mass energy and then evaluated for \five. This energy interpolation procedure, described in Ref.~\cite{LHCb-CONF-2014-003}, is employed to estimate the integrated and the \pt-dependent differential cross sections.
For the rapidity-differential cross sections, a further interpolation is performed as in the note~\cite{LHCb-CONF-2013-013}.
The rapidity dependence of d$\sigma_{\Upsilon\tomumu}^{\rm pp}/$d$y$ is fitted and integrated over the ranges matching the present Pb--Pb analysis.
The inclusion of CMS measurements~\cite{Sirunyan:2018nsz} constrains the curvature of the fit function down to $y=0$. 
The corresponding uncertainties are calculated as the quadratic sums of the uncertainties of the data propagated through the interpolation procedures and the spread of the results from the fitted functions.
The interpolation results are in agreement with the measured cross sections~\cite{ALICE:2021qlw} with systematic uncertainties smaller by more than a factor 2.

\subsection{Systematic uncertainties} \label{subsec:syst} 

This section describes the different sources of systematic uncertainties to compute the yields and the nuclear modification factors.
The raw yields as well as the \upstwos-to-\upsones yield ratio are evaluated as the average of the results obtained from the following fit variations.
The parameters of the Crystal~Ball distributions are estimated from MC simulations using either the \textsc{Geant3}~\cite{geant3} or \textsc{Geant4}~\cite{geant4} transport package. A set of tail parameters from simulations of pp~collisions at \five is also used for the signal extraction.
The ratios of widths characterising the signal shapes of \upstwos and \upsthrees are varied from 1 to 1.08 and from 1 to 1.14, respectively, in order to account for discrepancies between data and simulation. 
Several empirical functions are used to model the background shape, whether the event-mixing technique is applied or not. The raw spectrum is fitted with the sum of two decreasing exponentials as well as a pseudo-Gaussian function whose width varies linearly with mass. These functions are defined in the note~\cite{ALICEpage}. Once the mixed-event background is subtracted from the raw distribution, a single exponential or a power-law function are employed.
The fits are alternatively performed within two mass ranges, [6--13] and [7--14]~\GeVmass, to cover different invariant mass regions.
The final statistical uncertainty is calculated as the linear average over the uncertainty returned by the fit whereas the systematic uncertainty is estimated as one standard deviation of the distribution of the results. 
The systematic uncertainty for the signal extraction arises predominantly from the background description uncertainty.
At maximum, the relative systematic uncertainty ranges from $3.6\%$ to $10\%$ for the \upsones and from $9.7\%$ and $39\%$ for the \upstwos as a function of rapidity.
The signal extraction uncertainties are uncorrelated to any kinematic variable.

The uncertainties of the calculations of the \Axe corrections have multiple origins. The \pt and $y$ distributions in the initial MC conditions are modified by nuclear PDFs or by not considering any shadowing effect. The uncertainty associated to the choice of nPDF set is estimated as the maximum relative difference across the \Axe values obtained when switching the input shapes. The uncertainties due to the tracking, trigger and matching efficiencies are treated as in Ref.~\cite{Adam:2016rdg}, the dominant source being the dimuon reconstruction efficiency. Discrepancies between data and MC studies are propagated from single muon tracks to $\Upsilon$~mesons by multiplying the uncertainties with a factor two.
The uncertainties of the MC input shapes as well as of the tracking, trigger and matching efficiencies are uncorrelated with \pt and rapidity and fully correlated with the centrality. In addition, the tracking systematic uncertainty contains an uncorrelated component as a function of the centrality. The latter is negligible in the peripheral collisions and increases up to $1\%$ for the most central events.

Furthermore, the yields can be sensitive to the uncertainty associated to the definition of the centrality intervals.
The corresponding uncertainties are taken from a previous \jpsi analysis~\cite{Adam:2016rdg} as the statistical fluctuations are too large to make a sensitive estimate with the present $\Upsilon$~measurement. This uncertainty is negligible for the 0--90$\%$ interval.
The uncertainty assigned to the \TAA calculation is detailed in the references~\cite{Adam:2015ptt,ALICE-PUBLIC-2018-011}.
The number of equivalent minimum~bias events is obtained from the number of analysed dimuon-triggered events normalised by the probability of having a dimuon trigger in a MB event. Its uncertainty is estimated in Ref.~\cite{Acharya:2020puh} and is correlated with all kinematic variables. The uncertainties of the production cross sections in pp collisions are fully correlated with the centrality and uncorrelated as a function of \pt and rapidity.

All the sources of systematic uncertainties entering in the computation of the nuclear modification factor are summarised in Tables~\ref{tab:syst_summary1S} and~\ref{tab:syst_summary2S} for the \upsones and \upstwos, respectively. The ranges indicate the minimum and maximum relative uncertainties as a function of the given kinematic variable.
The uncertainties are common to the normalised yields. For the ratio of yields, only the uncertainty attributed to the signal extraction as well as the branching ratio uncertainties~\cite{PDG} are assumed not to cancel.

\begin{table}[h]
    \centering
    \caption{Summary of the relative systematic uncertainties (in $\%$) of the \raa for \upsones. The symbol ``$\oplus$" denotes the quadratic sum of a correlated and an uncorrelated component. The first value represents the component correlated with the column variable while the second value is the uncorrelated component.}
    
    \begin{tabular}{l|c|c|c|c}
        \hline
        Source of uncertainty & Integrated & Centrality & \pt & $y$ \\
        \hline
        Signal extraction & 4.4 & 3.9--7.4 & 3.6--9.1 & 3.6--10.2\\
        Monte Carlo input & 0.4 & 0.4 & 0.1--0.6 & 0.2--0.4  \\
        Tracking efficiency & 3 & $3~\oplus$ (0--1) & $1 \oplus 3$ & $1 \oplus 3$ \\
        Trigger efficiency & 3 & 3 & 0.5--2.6 & 1.5--4.0\\
        Matching efficiency & 1 & 1 & 1 & 1 \\
        Number of MB events & 0.5 & 0.5 & 0.5 & 0.5 \\
        \TAA & 1 & 0.7--2.4 & 1 & 1\\
        Centrality determination & - & 0.1--5.5 & - & - \\
        $\sigma_{\upsones\tomumu}^{\text{pp}}$ & 5.3 & 5.3 & 5.7--15.8 & 4.9--14.3\\
        \hline
    \end{tabular}
    \label{tab:syst_summary1S}
\end{table}

\begin{table}[h]
    \centering
    \caption{Summary of the relative systematic uncertainties (in $\%$) of the \raa for \upstwos. The symbol ``$\oplus$" denotes the quadratic sum of a correlated and an uncorrelated component. The first value represents the component correlated with the column variable while the second value is the uncorrelated component.}
    
    \begin{tabular}{l|c|c|c}
        \hline
        Source of uncertainty & Integrated & Centrality & $y$\\
        \hline
        Signal extraction & 18.4 & 14.0--23.4 & 9.7--39.0\\
        Monte Carlo input & 0.3 & 0.3 & $0.7$\\
        Tracking efficiency & 3 & $3~\oplus$ (0--1) & $1 \oplus 3$\\
        Trigger efficiency& 3 & 3 & 1.4--3.7\\
        Matching efficiency & 1 & 1 & 1\\
        Number of MB events & 0.5 & 0.5 & 0.5\\
        \TAA & 1 & 0.8--2.6 & 1\\
        Centrality determination & - & 0.1--0.3 & - \\
        $\sigma_{\upstwos\tomumu}^{\text{pp}}$ & 7.5 & 7.5 & 8.1--14.4\\
        \hline
    \end{tabular}
    
    \label{tab:syst_summary2S}
\end{table}

\section{Results and discussion}

\begin{figure}[h]
    \begin{center}
    \includegraphics[width=.49\linewidth]{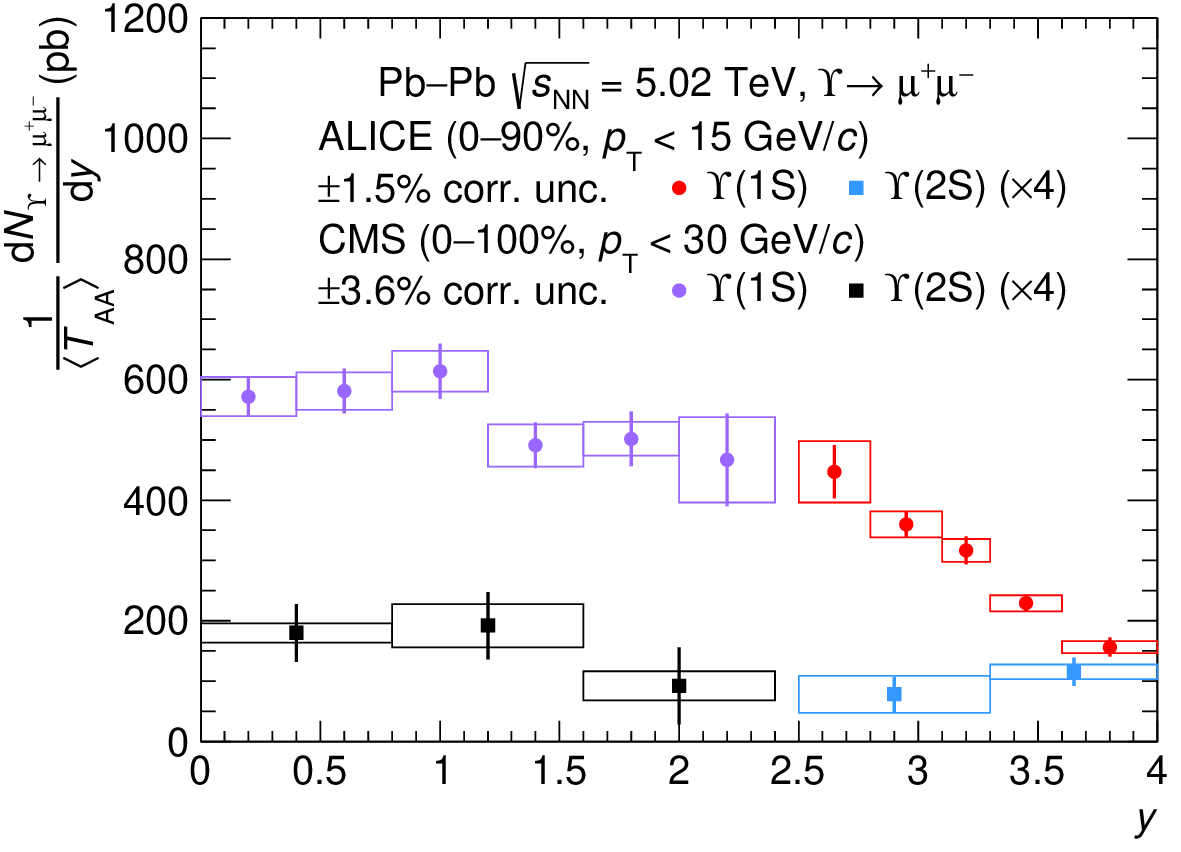}
    ~
    \includegraphics[width=.49\linewidth]{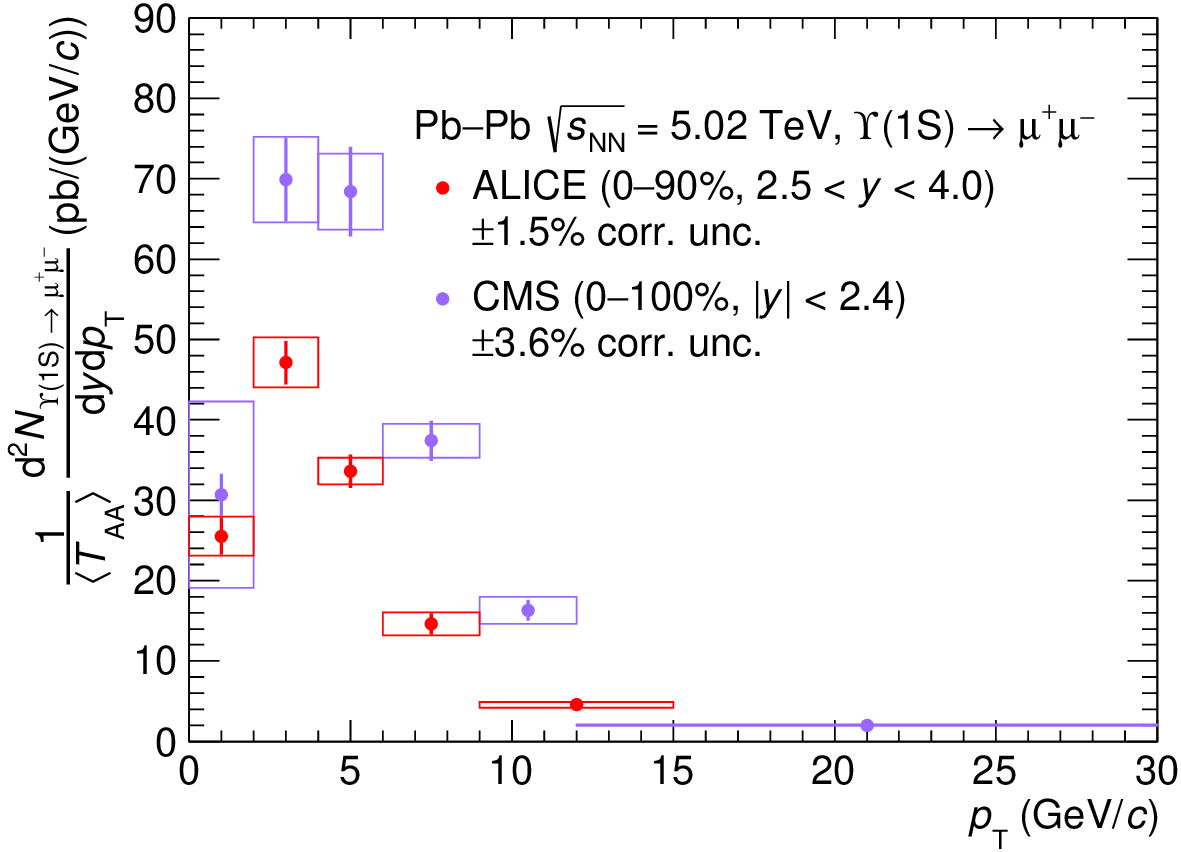}
    
    \caption{Rapidity (left) and \pt (right) differential measurements of normalised yields for inclusive $\Upsilon$~production in \PbPb collisions at \fivenn together with CMS measurements~\cite{Sirunyan:2018nsz}. The \upstwos results are multiplied by a factor 4 for better visibility. The global uncertainties of the nuclear overlap functions and of the number of MB events are not represented.} 
    \label{fig:yields}
    \end{center}
\end{figure}

The rapidity and \pt-differential yields of inclusive $\Upsilon$ production in \PbPb collisions, normalised by $\TAA=6.28\pm0.06$~mb$^{-1}$ for the 0--90$\%$ centrality interval~\cite{ALICE-PUBLIC-2018-011}, are presented in Fig.~\ref{fig:yields}.
The results are shown together with the CMS measurements performed at midrapidity~\cite{Sirunyan:2018nsz}. The vertical error bars represent the statistical uncertainties whereas the boxes correspond to the uncorrelated systematic uncertainties described in the previous section. 
Henceforth, this convention is adopted for all figures. The rapidity dependence indicates that the forward rapidity measurement spans an interesting dynamic range where the signal falls off significantly with respect to the approximate plateau reached around  midrapidity. For the first time, a significant \upstwos signal is observed in \PbPb collisions at forward rapidity. 
The measured $\pt$~spectrum of \upsones within the ALICE forward acceptance is softer than at midrapidity as also observed in pp and p--Pb~collisions~\cite{Andronic:2015wma}.

\begin{figure}[h]
    \begin{center}
    \includegraphics[width=.49\linewidth]{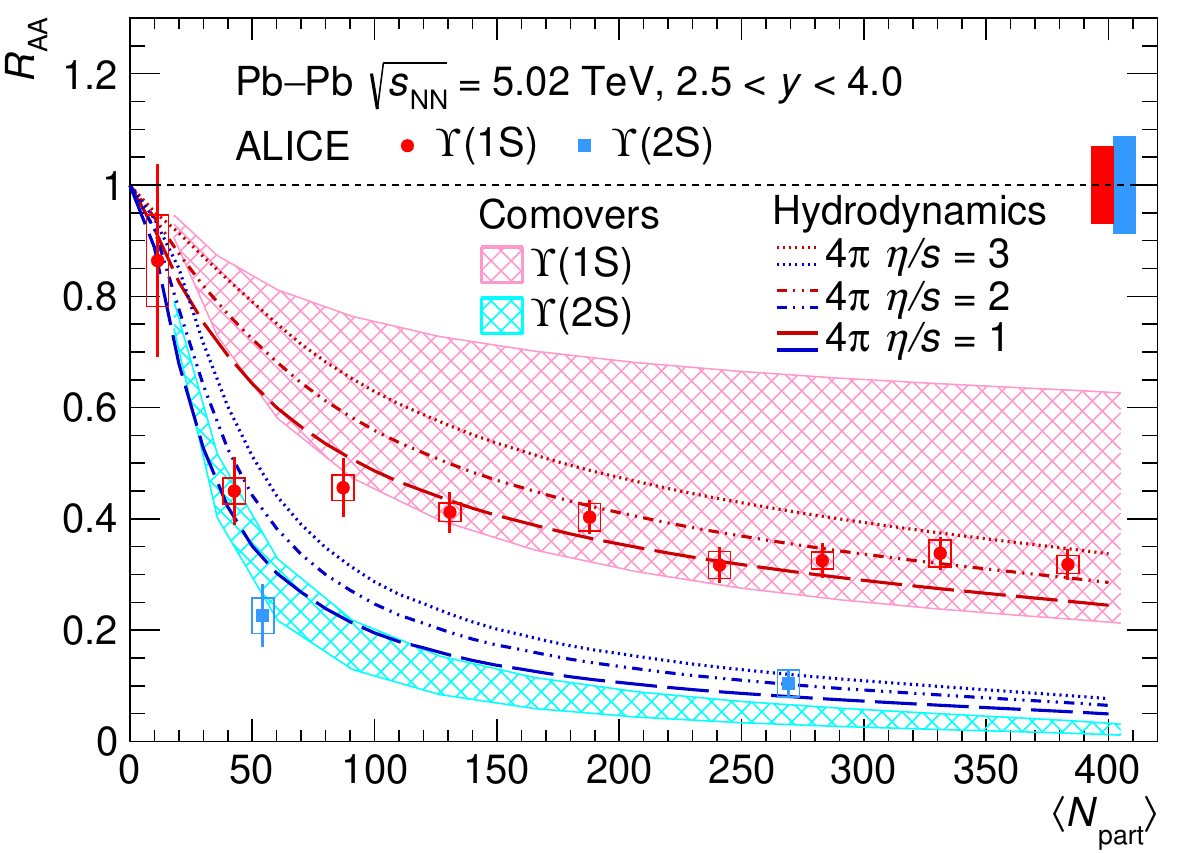}
    ~
    \includegraphics[width=.49\linewidth]{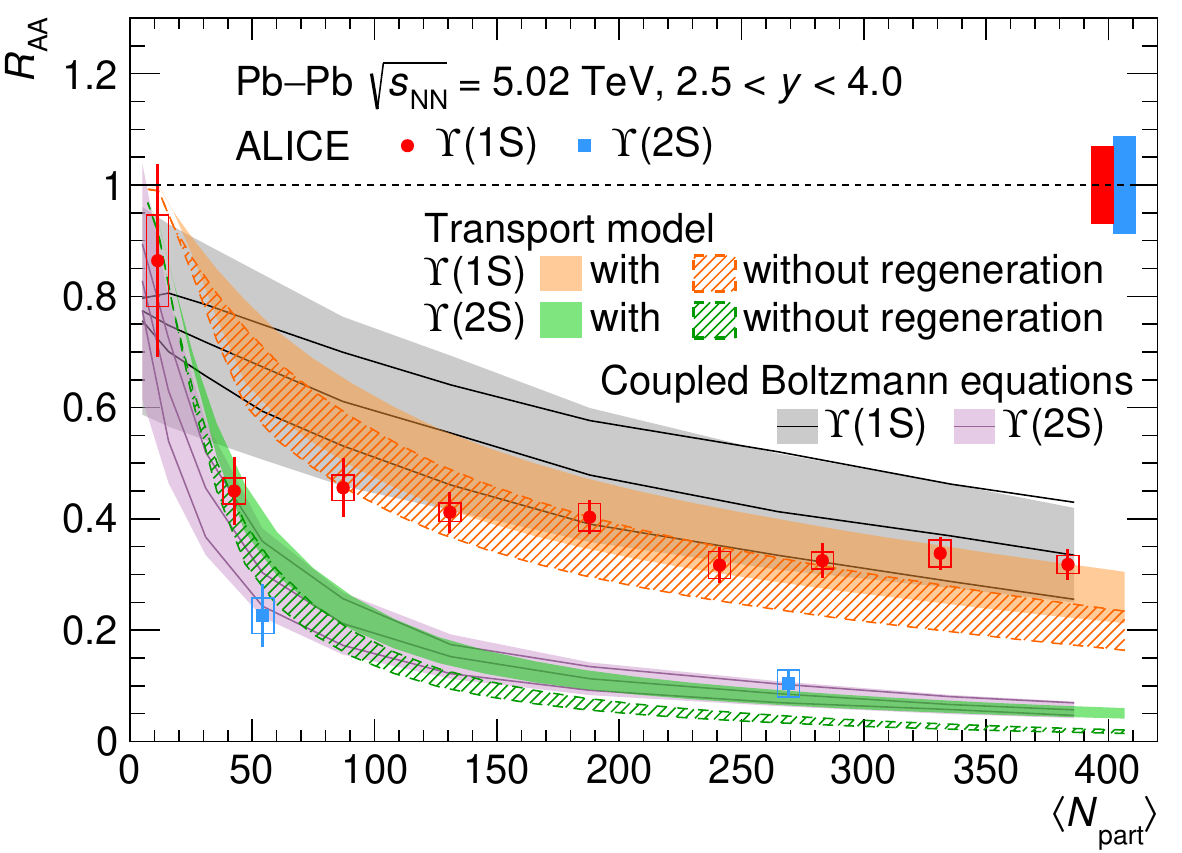}
    \caption{Nuclear modification factor of \upsones and \upstwos as a function of the average number of participants. The filled boxes at unity correspond to the relative uncertainties correlated with centrality. The results are compared with calculations from the comover and the hydrodynamic models~\cite{Ferreiro:2018wbd,Krouppa:2016jcl} in the left panel and with the transport descriptions~\cite{Du:2017qkv,Yao:2020xzw} in the right panel.
    }
    \label{fig:raa_npart}
    \end{center}
\end{figure}

The integrated nuclear modification factor is $0.353\pm0.012$~(stat.)~$\pm\, 0.029$~(syst.) for \upsones and $0.128\pm0.024$~(stat.)~$\pm\, 0.026$~(syst.) for \upstwos.
The dependence with the collision centrality is depicted in Fig.~\ref{fig:raa_npart} by representing the \raa as a function of the average number of participant nucleons $\langle \Npart \rangle$. The suppression of \upsones production with respect to pp collisions gets stronger towards more central collisions as reported by the CMS collaboration at midrapidity~\cite{Sirunyan:2018nsz}. The \raa of \upsones is compatible with unity in the 70--90$\%$ most peripheral interval and decreases to a plateau value of about~0.32 for $\langle \Npart \rangle > 200$. The nuclear modification factor of \upstwos is smaller by about a factor 2--3. The sizeable uncertainties of the measurements within the 0--30$\%$ and 30--90$\%$ centrality classes prevent any conclusion on a centrality dependence for \upstwos.

The measurements are compared with calculations based on transport and rate equations. 
Within the comover picture, quarkonia are dissociated via the interaction with surrounding particles in the final state~\cite{Ferreiro:2018wbd}. The revisited version of this model aims to explain the suppression of bottomonium production in both p--Pb and \PbPb collisions with the same assumptions. It takes into account the nuclear modification of parton distribution functions~(nPDFs). Uncertainties from the nCTEQ15 shadowing~\cite{Kusina:2017gkz} and the comover-$\Upsilon$ interaction cross sections are depicted together in the figures as grids. 
Predictions are also derived from the thermal modification of a complex heavy-quark potential inside an anisotropic plasma~\cite{Krouppa:2016jcl}. The survival probability of bottomonia is evaluated based on the local energy density, integrating the rate equation over the proper time of each state. 
The background medium is described with viscous hydrodynamics for three values of the shear viscosity-to-entropy density ratio $\eta/s$. These calculations do not include any modification of nuclear PDFs or any regeneration phenomenon.
The transport approaches describe an interplay of dissociation and regeneration mechanisms regulating the production of bottomonia at the QGP stage.
For the transport model~\cite{Du:2017qkv}, the medium evolves as an expanding isotropic fireball. 
Results are provided with and without the presence of a regeneration component. The width of the bands represents the modification of the PDF modelled by an effective scale factor on the initial number of b$\overline{\rm b}$~pairs.
This scale factor is neglected in peripheral collisions and is varied between no modification up to $30\%$ suppression at forward rapidity in the most central collisions.
In the framework of coupled Boltzmann equations~\cite{Yao:2020xzw}, the regeneration is dominated by real-time recombinations of correlated heavy-quark pairs. 
The simulation of the collision system includes the EPPS16 nPDF parametrisation~\cite{Eskola:2016oht}. 
In the figures, the calculations are shown with a band due to the nPDF uncertainty and with three curves from the variation of the coupling constants.

Inclusive production can be decomposed into the direct production component and the production from feed-down of higher bottomonium resonances.
Direct \upsones production constitutes approximately $70\%$ to the total inclusive cross section in pp~collisions at the LHC~\cite{Andronic:2015wma}. The feed-down contributions of \mbox{P-wave} states to excited $\Upsilon$~production are not measured at low transverse momentum and are extrapolated. All models treat the excited states and their decay chains to $\Upsilon$~mesons. 
They account for a feed-down contribution to $\upsones$~production in pp~collisions consistent with the measured values, but with varying assumptions for the feed-down of the excited states. An uncertainty based on the variation of the feed-down contributions is only considered for the comover interaction model~\cite{Ferreiro:2018wbd}.

The predictions from the comover and the hydrodynamic models are compared with the data in the left panel of Fig.~\ref{fig:raa_npart} while the calculations from the transport approaches are shown in the right panel of the same figure.
The various calculations reproduce the trend of the data within uncertainties. For the \upsones, the measurement points lie on the lower limit of the comover interaction model~\cite{Ferreiro:2018wbd} and of the coupled Boltzmann equations~\cite{Yao:2020xzw}. The sharp slope expected in all cases for the \raa of \upstwos is not measurable because of statistical limitations.
The current models do not account for the spatial dependence of the nPDF modification: stronger effects are expected for nuclei probed close to their centre. The impact on measurements has been discussed early on~\cite{Emelyanov:1999pkc}, a more recent extraction can be found in Ref.~\cite{Helenius:2012wd}. Future models should consider this phenomenon in particular for the discussion of centrality-dependent studies.

\begin{figure}[h]
    \begin{center}
    \includegraphics[width=.49\linewidth]{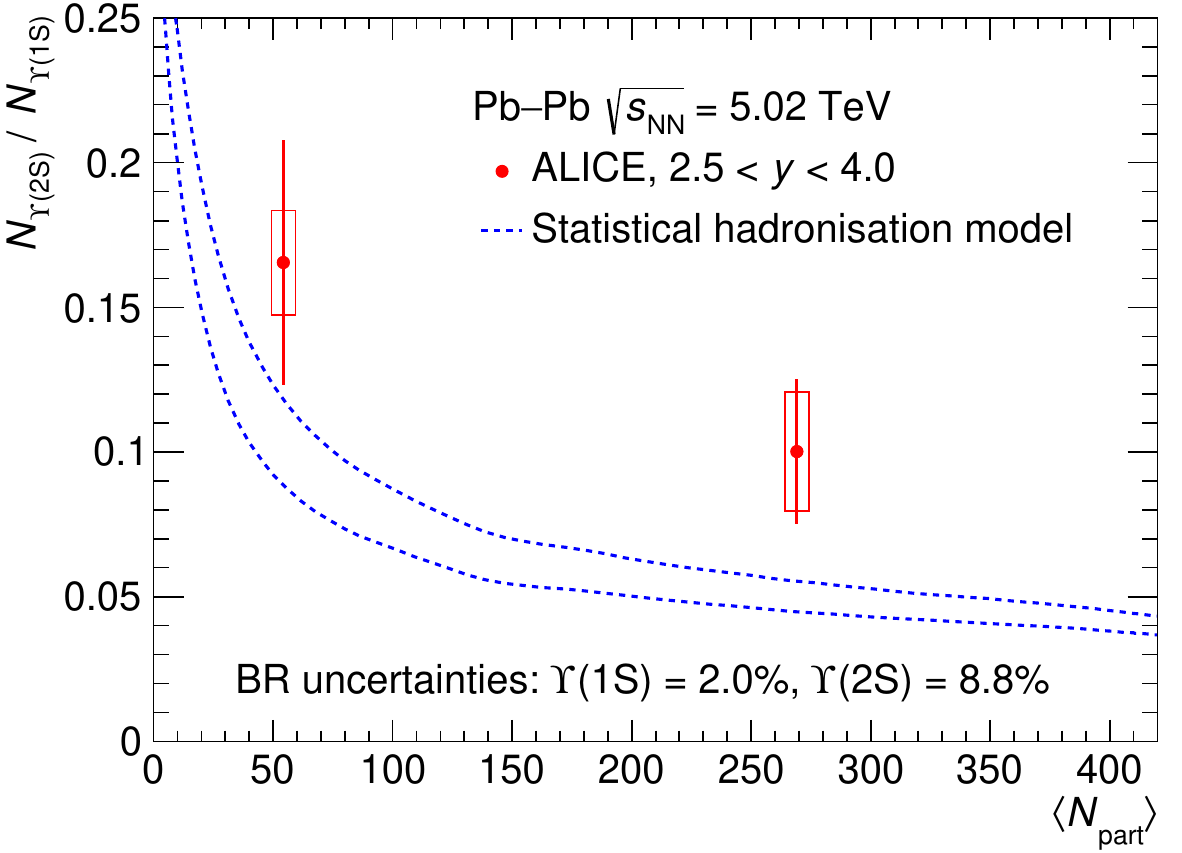}
    ~
    \includegraphics[width=.49\linewidth]{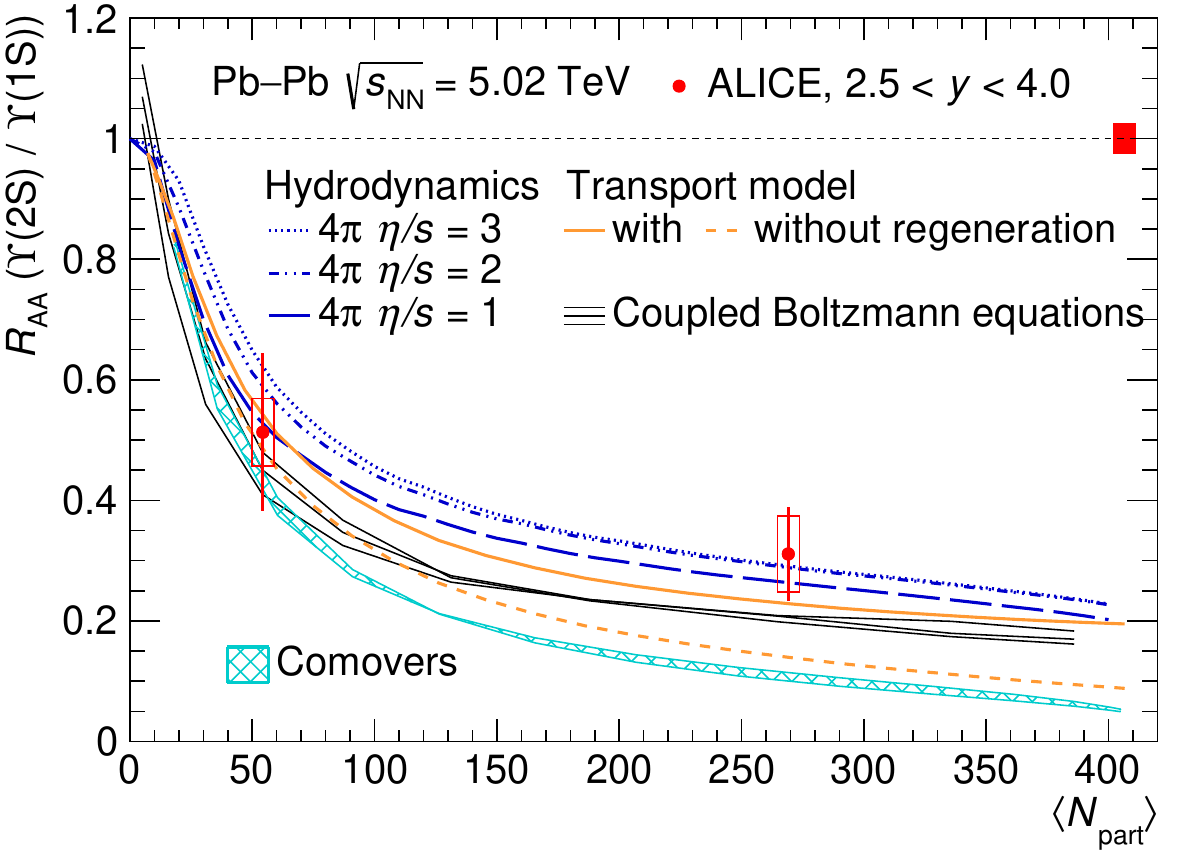}
    \caption{(Left) Ratio of \upstwos-to-\upsones yields as a function of the average number of participants. The results are displayed on top of the statistical hadronisation model values~\cite{Andronic:2017pug}. The two curves represent the uncertainty of the pp-like contribution of the corona of the nuclear overlap.
    (Right) Relative nuclear modification factor along with the predictions from the comover interaction model~\cite{Ferreiro:2018wbd}, hydrodynamic calculations~\cite{Krouppa:2016jcl}, from the transport model~\cite{Du:2017qkv} and calculations based on the coupled Boltzmann equations~\cite{Yao:2020xzw}. The filled red box at unity denotes the uncertainty on the \upstwos-to-\upsones cross section ratio in pp collisions.
    }
    \label{fig:ratio}
    \end{center}
\end{figure}

Before studying the different approaches via the relative suppression, the  \upstwos-to-\upsones yield ratio in \PbPb~collisions shown in the left panel of Fig.~\ref{fig:ratio} is compared with the statistical hadronisation model~\cite{Andronic:2017pug}.
It assumes that the final state light-flavour hadron yields are calculable from hadron resonance gas densities with a common chemical freeze-out temperature and the baryochemical potential tracing closely the phase boundary from the QGP to hadrons.
The approach has been extended to heavy-flavour production assuming a kinetic equilibration of the heavy quarks prior to freeze-out and total heavy quark conservation using the production from initial hard scatterings as an input. The model calculates the abundances of various heavy-flavour species assuming thermal weights. For non-central collisions, a contribution from pp-like scattering behaviour at the surface of the interaction zone is introduced.
The calculation underestimates the measured \upstwos-to-\upsones yield ratio for the 0--30$\%$ most central collisions. Taking into account all the uncertainties, the deviation is about one sigma.
Comparisons with other measurements in the bottom sector  are required to further test the applicability of the model.

The relative nuclear modification factor defined by Eq.~\ref{eq:double_ratio} is an appropriate observable to confront the different approaches. 
Considerations of effects common to both states are thus expected to disappear as indicated by the smaller uncertainties in the right panel of Fig.~\ref{fig:ratio} compared to Fig.~\ref{fig:raa_npart}. The relative uncorrelated systematic uncertainties are $4\%$ smaller when accounting for correlations in the signal extraction, while the systematic uncertainty correlated with the centrality is reduced from $9.2\%$ to $2.5\%$.
Integrated over the 0--90$\%$ class, the \upstwos-to-\upsones \raa is $0.360\pm0.069$~(stat.)~$\pm\, 0.055$~(syst.) i.e.\ $7.2\sigma$ from unity.
The comover interaction model shows a deviation with respect to the measurement for the 0--30$\%$ most central collisions, also noticed in a comparison with CMS data~\cite{Ferreiro:2018wbd}.
The hydrodynamic calculations~\cite{Krouppa:2016jcl}  describe well the data within the present experimental uncertainties.
Interestingly, within the transport approaches~\cite{Du:2017qkv,Yao:2020xzw}, larger double ratio values are achieved by the large regeneration component of the \upstwos~production.
With more precise measurements, the relative \raa could serve as a model discriminator thanks to the cancellation of sources of uncertainty and the different slopes between the models.

\begin{figure}[h]
    \begin{center}
    \includegraphics[width=.5\linewidth]{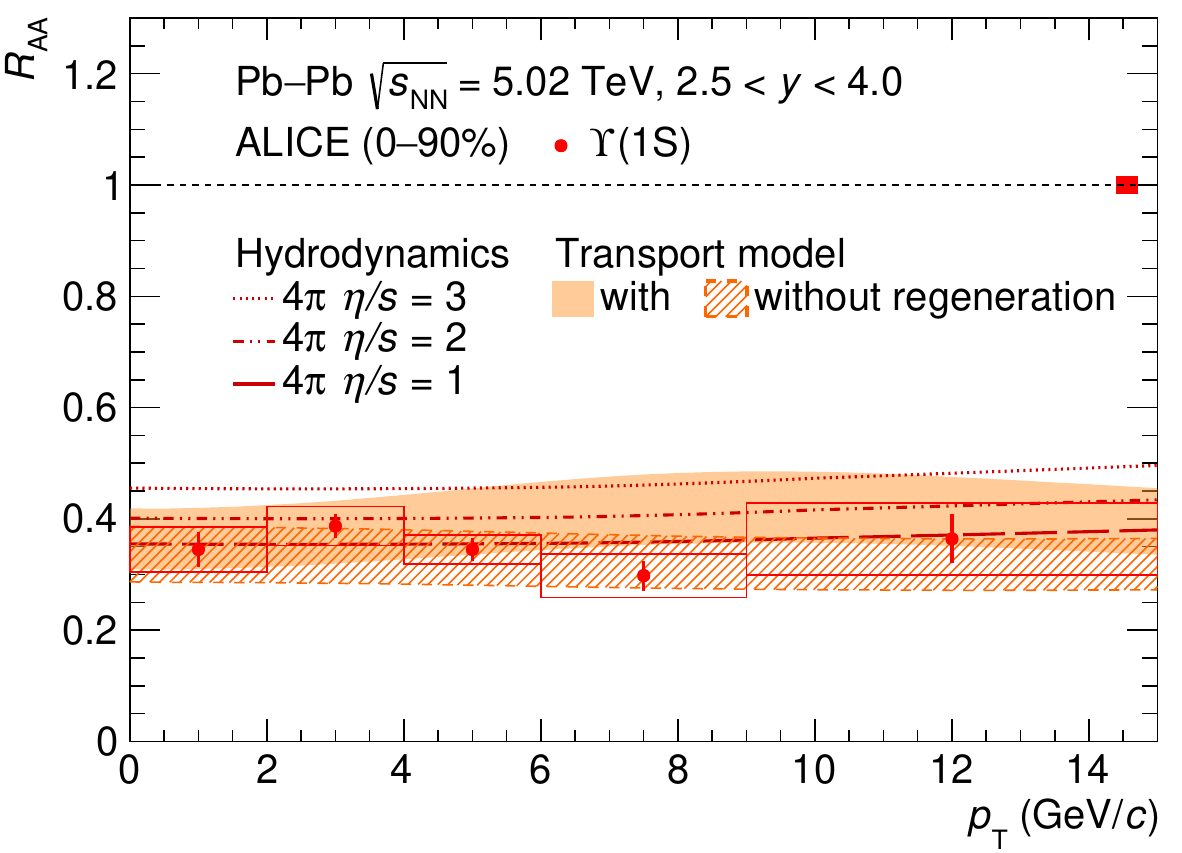}
    \caption{Nuclear modification factor of \upsones as a function of the transverse momentum. The red box at unity corresponds to the global uncertainty correlated with \pt.
    Predictions from the hydrodynamic~\cite{Krouppa:2016jcl} and transport~\cite{Du:2017qkv} models are also shown.}
    \label{fig:raa_pt}
    \end{center}
\end{figure}

In the following, differential studies of the nuclear modification factor are carried out to scrutinise the suppression features.
The dependence of the \raa on the transverse momentum is investigated in Fig.~\ref{fig:raa_pt} for the 0--90$\%$ centrality interval. 
No significant variation is observed up to 15~\GeVc, in line with hydrodynamic and transport model expectations. The present \upsones~measurement disfavours the hydrodynamic calculation for the highest shear viscosity-to-entropy density ratio.
It is interesting to note that the nuclear modification factor in p--Pb collisions exhibits a significant \pt dependence in both forward and backward regions~\cite{Aaij:2018scz,Acharya:2019lqc}.
The difference in spectral shape between proton--nucleus and nucleus--nucleus collisions should receive a close attention from a phenomenological point of view.

\begin{figure}[h]
    \begin{center}
    \includegraphics[width=.49\linewidth]{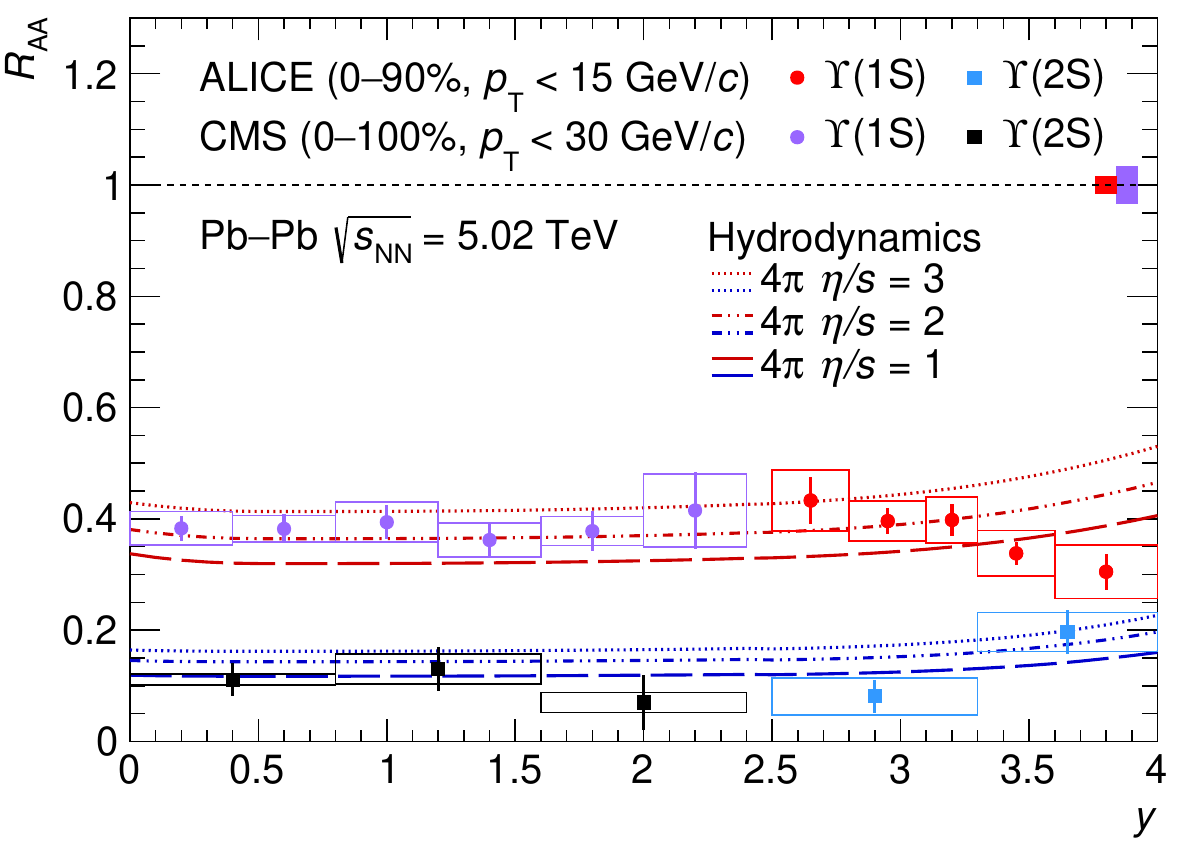}
    ~
    \includegraphics[width=.49\linewidth]{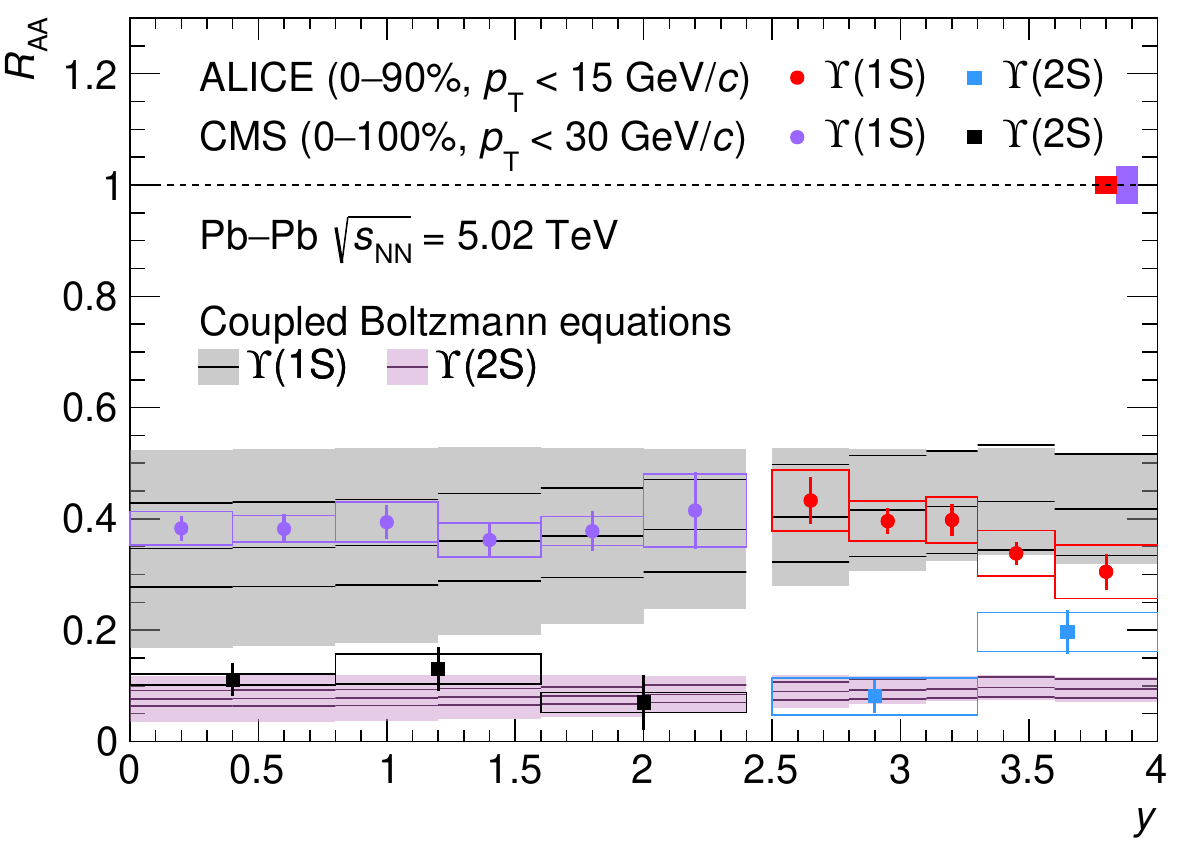}
    \caption{Nuclear modification factor of \upsones and \upstwos as a function of rapidity. The red and violet filled boxes at unity correspond to the global uncertainties common to both $\Upsilon$~states from the ALICE and CMS measurements. The experimental data are compared with hydrodynamic~\cite{Krouppa:2016jcl} and coupled Boltzmann equations~\cite{Yao:2020xzw} calculations on the left and right panel, respectively.}
    \label{fig:rapidityRAA}
    \end{center}
\end{figure}

Figure~\ref{fig:rapidityRAA} shows the rapidity dependence of the nuclear modification factors measured by ALICE and CMS~\cite{Sirunyan:2018nsz}, and the results are compared with the hydrodynamic model as well as with the calculations based on the coupled Boltzmann transport equations.
The experimental data indicate a plateau value of the \upsones nuclear modification factor of around 0.4 between midrapidity and $y\approx 3$.  The two most forward measurement points hint of a decrease of the nuclear modification factor down to a value close to 0.3: the \raa is lower by $2\sigma$ in the most forward rapidity interval with respect to the central range of the ALICE measurement.
The hydrodynamic calculations indicate the opposite behaviour. In this model, the rapidity profile inherits from the initial conditions of the simulated medium~\cite{Krouppa:2016jcl}. 
The results from the coupled Boltzmann equations exhibit a structure induced by the rapidity-dependent impact of the used nPDF~\cite{Yao:2020xzw}. The curves cannot describe the CMS and ALICE measurement consistently, albeit the most forward data points lie on the limit of the nPDF uncertainty band.
These discrepancies may point towards a physical mechanism not captured in the presently available models.
This behaviour will need to be scrutinised further in future analyses improving the current experimental and theoretical uncertainties.
Interestingly, the nuclear modification factor of \upsones calculated with the coherent energy loss model shows a decreasing trend towards forward rapidity at \twosevensixnn~\cite{Arleo:2014oha}, even though the model does not reproduce the overall suppression at this centre-of-mass energy.
This different behaviour is caused by the conjunction of a non-constant rapidity distribution and the rapidity shift due to the heavy-quark pair medium interaction. All other models do not consider the latter effect.
The \raa of the \upstwos is independent of the rapidity within uncertainties in the measured interval with values between 0.07 and 0.20. The rapidity dependence of the models for \upstwos is similar to the one observed for the ground state and is compatible with the experimental measurements.

\section{Conclusions}

The results presented in this article provide a detailed measurement of the \upsones~production as well as the first significant measurement of the \upstwos state at forward rapidity in \PbPb collisions at the LHC. For the \mbox{0--90$\%$} centrality class, the nuclear modification factor is $0.353\pm0.012$~(stat.)~$ \pm\, 0.029$~(syst.) for \upsones and $0.128\pm0.024$~(stat.)~$\pm\, 0.026$~(syst.) for \upstwos. The corresponding excited-to-ground state ratios are in agreement with hydrodynamic calculations~\cite{Krouppa:2016jcl}, a transport model with a regeneration component~\cite{Du:2017qkv}, and predictions from coupled Boltzmann equations~\cite{Yao:2020xzw}. The results are in tension with the comover interaction model~\cite{Ferreiro:2018wbd} and with the statistical hadronisation model~\cite{Andronic:2017pug} for the 0--30$\%$ most central collisions.
Taken together with the CMS data, these measurements constrain the rapidity dependence of $\Upsilon$~suppression with respect to proton--proton collisions. 
The available models describing bottomonium production in heavy-ion collisions do not capture the rapidity dependence  observed for the \raa of \upsones in the ALICE acceptance.

Bottomonia are privileged observables to understand the formation and the interaction of bound states in strongly interacting matter and hence to learn about the degrees of freedom of the QGP. This measurement is one of the starting points for more differential studies of bottomonium production. Flow, $\Upsilon$(3S)~measurements as well as better constraints on parton densities in nuclear collisions, feed-down chains and beauty production cross sections will become available in the upcoming years at the LHC~\cite{Citron:2018lsq}.


\newenvironment{acknowledgement}{\relax}{\relax}
\begin{acknowledgement}
\section*{Acknowledgements}
We thank the phenomenology groups for their predictions. The authors would like to extend special thanks to Xiaojian Du, Elena Gonzalez Ferreiro and Xiaojun Yao for enlightening discussions.


The ALICE Collaboration would like to thank all its engineers and technicians for their invaluable contributions to the construction of the experiment and the CERN accelerator teams for the outstanding performance of the LHC complex.
The ALICE Collaboration gratefully acknowledges the resources and support provided by all Grid centres and the Worldwide LHC Computing Grid (WLCG) collaboration.
The ALICE Collaboration acknowledges the following funding agencies for their support in building and running the ALICE detector:
A. I. Alikhanyan National Science Laboratory (Yerevan Physics Institute) Foundation (ANSL), State Committee of Science and World Federation of Scientists (WFS), Armenia;
Austrian Academy of Sciences, Austrian Science Fund (FWF): [M 2467-N36] and Nationalstiftung f\"{u}r Forschung, Technologie und Entwicklung, Austria;
Ministry of Communications and High Technologies, National Nuclear Research Center, Azerbaijan;
Conselho Nacional de Desenvolvimento Cient\'{\i}fico e Tecnol\'{o}gico (CNPq), Financiadora de Estudos e Projetos (Finep), Funda\c{c}\~{a}o de Amparo \`{a} Pesquisa do Estado de S\~{a}o Paulo (FAPESP) and Universidade Federal do Rio Grande do Sul (UFRGS), Brazil;
Ministry of Education of China (MOEC) , Ministry of Science \& Technology of China (MSTC) and National Natural Science Foundation of China (NSFC), China;
Ministry of Science and Education and Croatian Science Foundation, Croatia;
Centro de Aplicaciones Tecnol\'{o}gicas y Desarrollo Nuclear (CEADEN), Cubaenerg\'{\i}a, Cuba;
Ministry of Education, Youth and Sports of the Czech Republic, Czech Republic;
The Danish Council for Independent Research | Natural Sciences, the VILLUM FONDEN and Danish National Research Foundation (DNRF), Denmark;
Helsinki Institute of Physics (HIP), Finland;
Commissariat \`{a} l'Energie Atomique (CEA) and Institut National de Physique Nucl\'{e}aire et de Physique des Particules (IN2P3) and Centre National de la Recherche Scientifique (CNRS), France;
Bundesministerium f\"{u}r Bildung und Forschung (BMBF) and GSI Helmholtzzentrum f\"{u}r Schwerionenforschung GmbH, Germany;
General Secretariat for Research and Technology, Ministry of Education, Research and Religions, Greece;
National Research, Development and Innovation Office, Hungary;
Department of Atomic Energy Government of India (DAE), Department of Science and Technology, Government of India (DST), University Grants Commission, Government of India (UGC) and Council of Scientific and Industrial Research (CSIR), India;
Indonesian Institute of Science, Indonesia;
Istituto Nazionale di Fisica Nucleare (INFN), Italy;
Institute for Innovative Science and Technology , Nagasaki Institute of Applied Science (IIST), Japanese Ministry of Education, Culture, Sports, Science and Technology (MEXT) and Japan Society for the Promotion of Science (JSPS) KAKENHI, Japan;
Consejo Nacional de Ciencia (CONACYT) y Tecnolog\'{i}a, through Fondo de Cooperaci\'{o}n Internacional en Ciencia y Tecnolog\'{i}a (FONCICYT) and Direcci\'{o}n General de Asuntos del Personal Academico (DGAPA), Mexico;
Nederlandse Organisatie voor Wetenschappelijk Onderzoek (NWO), Netherlands;
The Research Council of Norway, Norway;
Commission on Science and Technology for Sustainable Development in the South (COMSATS), Pakistan;
Pontificia Universidad Cat\'{o}lica del Per\'{u}, Peru;
Ministry of Science and Higher Education, National Science Centre and WUT ID-UB, Poland;
Korea Institute of Science and Technology Information and National Research Foundation of Korea (NRF), Republic of Korea;
Ministry of Education and Scientific Research, Institute of Atomic Physics and Ministry of Research and Innovation and Institute of Atomic Physics, Romania;
Joint Institute for Nuclear Research (JINR), Ministry of Education and Science of the Russian Federation, National Research Centre Kurchatov Institute, Russian Science Foundation and Russian Foundation for Basic Research, Russia;
Ministry of Education, Science, Research and Sport of the Slovak Republic, Slovakia;
National Research Foundation of South Africa, South Africa;
Swedish Research Council (VR) and Knut \& Alice Wallenberg Foundation (KAW), Sweden;
European Organization for Nuclear Research, Switzerland;
Suranaree University of Technology (SUT), National Science and Technology Development Agency (NSDTA) and Office of the Higher Education Commission under NRU project of Thailand, Thailand;
Turkish Atomic Energy Agency (TAEK), Turkey;
National Academy of  Sciences of Ukraine, Ukraine;
Science and Technology Facilities Council (STFC), United Kingdom;
National Science Foundation of the United States of America (NSF) and United States Department of Energy, Office of Nuclear Physics (DOE NP), United States of America.
\end{acknowledgement}

\bibliographystyle{utphys}   
\bibliography{bibliography}

\newpage
\appendix
%
%

\section{The ALICE Collaboration}
\label{app:collab}
\small
\begin{flushleft} 

\bigskip 

S.~Acharya$^{\rm 142}$, 
D.~Adamov\'{a}$^{\rm 97}$, 
A.~Adler$^{\rm 75}$, 
J.~Adolfsson$^{\rm 82}$, 
G.~Aglieri Rinella$^{\rm 35}$, 
M.~Agnello$^{\rm 31}$, 
N.~Agrawal$^{\rm 55}$, 
Z.~Ahammed$^{\rm 142}$, 
S.~Ahmad$^{\rm 16}$, 
S.U.~Ahn$^{\rm 77}$
% ...

\bigskip

\bigskip 

\textbf{\Large Affiliation Notes}

\bigskip 

% ...
$^{\rm VI}$ Also at: Institute of Theoretical Physics, University of Wroclaw, Poland\\

\bigskip

\bigskip 

\textbf{\Large Collaboration Institutes}

\bigskip 

% ...
$^{16}$ Department of Physics, Aligarh Muslim University, Aligarh, India\\
% ...
$^{31}$ Dipartimento DISAT del Politecnico and Sezione INFN, Turin, Italy\\
% ...
$^{35}$ European Organization for Nuclear Research (CERN), Geneva, Switzerland\\
% ...
$^{55}$ INFN, Sezione di Bologna, Bologna, Italy\\
% ...
$^{75}$ Johann-Wolfgang-Goethe Universit\"{a}t Frankfurt Institut f\"{u}r Informatik, Fachbereich Informatik und Mathematik, Frankfurt, Germany\\
% ...
$^{77}$ Korea Institute of Science and Technology Information, Daejeon, Republic of Korea\\
% ...
$^{82}$ Lund University Department of Physics, Division of Particle Physics, Lund, Sweden\\
% ...
$^{97}$ Nuclear Physics Institute of the Czech Academy of Sciences, \v{R}e\v{z} u Prahy, Czech Republic\\
% ...
$^{142}$ Variable Energy Cyclotron Centre, Homi Bhabha National Institute, Kolkata, India\\
% ...

\bigskip 

\end{flushleft} 
  
\end{document}